\documentclass[10pt]{article} 
\usepackage{amsfonts,amssymb,slashed,makeidx,latexsym,setspace}
\usepackage{graphicx,graphics,floatflt,amssymb,epsf,rotate,subfigure} 
\usepackage{times,cite,color,epsfig,epsf}
\usepackage{multirow}
\usepackage{hyperref}
\def\five{\Phi_5}
\def\fivebar{\Phi_{\bar 5}}
\def\three{\Sigma_3}
\def\eight{\Sigma_8}

\begin{document}

\begin{flushright}
SINP/TNP/2013/06, IPMU-13-0079
\end{flushright}

\begin{center}
{\Large \bf A natural scenario for heavy colored and light
  uncolored superpartners} \\
\vspace*{1cm} \renewcommand{\thefootnote}{\fnsymbol{footnote}} { {\sf
Gautam Bhattacharyya${}^{1}$},
{\sf Biplob Bhattacherjee${}^{2}$}, {\sf Tsutomu
  T. Yanagida${}^{2}$}, and {\sf Norimi Yokozaki${}^{2}$}
}
\\
\vspace{10pt} {\small ${}^{1)}$ {\em Saha Institute of Nuclear
    Physics, 1/AF Bidhan Nagar, Kolkata 700064, India} \\ ${}^{2)}$
  {\em Kavli IPMU, TODIAS, University of Tokyo, Kashiwa 277-8583,
    Japan} \\ } \normalsize
\end{center}

\begin{abstract}

Influenced by the current trend of experimental data, especially from
the LHC, we construct a supersymmetric scenario where a natural
dynamics makes the squarks and gluino super-heavy (order 10 TeV) while
keeping the sleptons and the weak gauginos light (100-500 GeV). The
dynamics relies on the interfusion of two underlying ideas: ($i$)
gauge mediation of supersymmetry breaking with two messenger
multiplets, one transforming as a triplet of weak SU(2) and the other
as an octet of color SU(3); ($ii$) perturbative gauge coupling
unification at the string scale even with these incomplete SU(5)
multiplets.  Interestingly, the relative magnitude of the triplet and
octet messenger scales that ensures gauge unification at the two-loop
level also helps to naturally keep the uncolored superpartners light
while making the colored ones heavy.

\end{abstract}

\setcounter{footnote}{0}
\renewcommand{\thefootnote}{\arabic{footnote}}


If the recently discovered scalar particle with a mass of around 125
GeV at the CERN Large Hadron Collider (LHC)
\cite{Aad:2012tfa,Chatrchyan:2012ufa} has to be identified with the
lightest supersymmetric (SUSY) Higgs boson then, within the framework
of the minimal supersymmetric standard model (MSSM), the stop squarks
are expected to be rather heavy (order 10 TeV) having a substantial
mixing between their left and right components
\cite{OYY,Okada:1990gg}. Side by side, the non-observation of the
first two generation squarks and the gluino in the 7 and 8 TeV run of
the LHC, with the lower limits on their masses now pushed to around
1.5 TeV \cite{susy_sq_glu}, sends us an early alert that they might
remain elusive even in the 14 TeV run of the LHC.  The absence of any
statistically significant indirect evidence of new physics in meson
oscillation and decays so far, measured with increasingly high
precision by the Belle, BaBar and LHCb Collaborations, also endorses
the view that colored superparticles might not lie within the
periphery of the LHC territory. On the other hand, the uncolored
superparticles, namely, the sleptons and the neutralinos/charginos,
are (and would remain) relatively less constrained by the LHC
\cite{susy_ew,CMS1}. Interestingly, the (3.3-3.6)-$\sigma$ deviation
of the measured $(g-2)$ of muon \cite{muon_g_exp} from its standard
model (SM) expectation \cite{g-2_hagiwara2011, g-2_davier2010} might
hint towards a light smuon and gaugino/higgsino \cite{susy_muon_g}.
Additionally, the reported excess of the diphoton events in Higgs
decay by the ATLAS Collaboration \cite{atlas_diphoton} can be
explained by the presence of light staus
\cite{Carena:2012gp,Cao:2012fz,Ajaib:2012eb} (though the diphoton
decay rate reported by the CMS Collaboration \cite{cms_diphoton} may
not be construed as an excess).  Even if these apparent discrepancies
eventually disappear, the possible existence of light sleptons and
weak neutralinos/charginos still merits a careful investigation
especially in view of precision measurements at the upcoming
International Linear Collider (ILC).

Given the present experimental situation as narrated above, what kind
of a broad picture can we draw about plausible supersymmetric models?
For example, can we conceive of a scenario that naturally accommodates
heavy colored (order 10 TeV) and light uncolored (order 100 GeV)
superparticles?  This question is very pertinent and timely as in many
scenarios, notably the gravity mediated supersymmetry breaking models,
the heaviness of squarks also implies a set of heavy sleptons ({\em
  modulo} gaugino induced splitting by renormalization group (RG)
running).  Gauge mediated supersymmetry breaking (GMSB) models
\cite{GMSB} offer a way-out by introducing messenger particles in the
intermediate scale, well below $M_G \simeq 2 \cdot 10^{16}$ GeV, and
at that scale the squark masses are generated being proportional to
the strong gauge coupling and slepton masses are generated being
proportional to the weak gauge coupling. Thus the masses of squarks
and sleptons are split right at the time of generation and the
relative separation between grows even further when those masses are
run down to the weak scale.  Note that in the minimal GMSB scenario
one employs a $\five$ and a $\fivebar$ messenger multiplets, which
transform as a fundamental ${\bf 5}$ and a ${\bf \bar{5}}$
representation of SU(5), respectively, for the generation of {\em all}
superparticle masses.  Still, it is difficult to keep sleptons too
light if the squarks become too heavy.

In this paper, we resurrect an old idea in the GMSB context which
introduces an unconventional choice of messenger particles
\cite{yanagida-old}. One of the key features of this scenario is that
the sources for squark/gluino and slepton/weak gaugino mass generation
are completely de-linked, which allows us to naturally maintain two
orders of mass splitting between them. Instead of taking $\five$ as
$\fivebar$, here we employ messenger multiplets transforming as an
adjoint octet ($\eight$) of color SU(3) and an adjoint triplet
($\three$) of weak SU(2) \cite{yanagida-old}. The choices are not
completely arbitrary as the origin of these states can be traced to
the non-Goldstone modes of the scalar adjoint ${\bf 24}$-plet of
SU(5).  The superpotential of the messenger sector reads
\begin{eqnarray}
\label{eq:mess1}
W_{\rm mess}= (M_8 + \lambda_8 X ) {\rm Tr}(\Sigma_8^2) + (M_3 +
\lambda_3 X ) {\rm Tr}(\Sigma_3^2) \, , 
\end{eqnarray}
where the $F$-term vacuum expectation value (vev) $F_X$ of the hidden
sector superfield $X$ transmits supersymmetry breaking to the
observable sector via the messenger multiplets\footnote{For a recent
  discussion with a complete {\bf 24}-plet messenger multiplet
  transforming in the adjoint of SU(5), see
  \cite{yanagida-recent}.}. The following consequences deserve special
attention:

\begin{floatingfigure}[r]{0.4\textwidth} 
\includegraphics[scale=0.95]{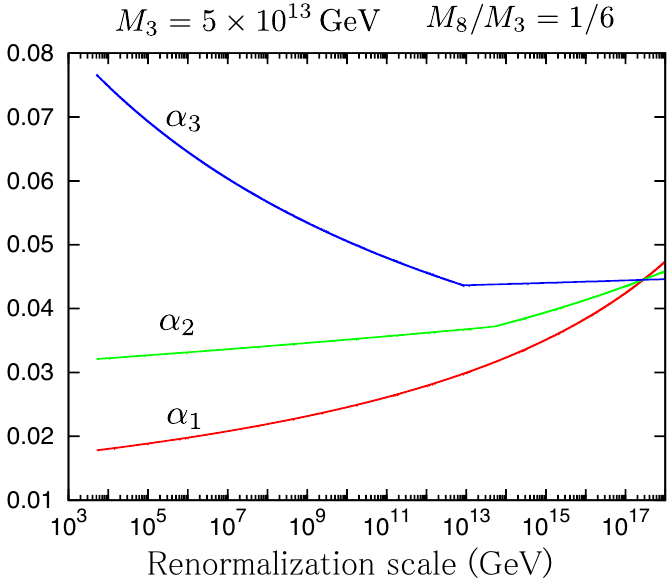}
\caption{\small Unification of the gauge couplings with SU(3) octet and SU(2)
  triplet messengers with their masses at around $10^{13}$
  GeV. Two-loop RG evolution has been used, and we have used
  $\alpha_S(M_Z) = 0.1184$ and $m_{\rm SUSY}= $ 5 TeV.}
\label{fig:gut}
\end{floatingfigure}

\noindent $(i)$~ Even in the absence of complete SU(5) multiplets, the
presence of an identical number of $\three ({\bf 1, 3, Y=0})$ and
$\eight ({\bf 8, 1, Y=0})$ messenger multiplets still ensures
perturbative gauge coupling unification at a scale somewhat higher
than $M_G$ \cite{yanagida-original}. More specifically, if the masses
of these states are around $10^{(13-14)}$ GeV, then unification occurs
even with these incomplete multiplets at around the string scale
$M_{\rm str} \approx 5 \cdot 10^{17}$ GeV, which is the scale where
the gravitational and gauge couplings are perturbatively unified
\cite{string}. This can be easily understood using the one-loop
beta-functions of the gauge couplings. The gauge couplings at the
string scale are given by
\begin{eqnarray}
\label{eq:rgemod}
\alpha_1^{-1} (M_{\rm str}) &=& \alpha_1^{-1} (m_{\rm SUSY}) -
\frac{b_1}{2\pi} \ln \frac{M_{\rm str}}{m_{\rm SUSY}}, \nonumber
\\ \alpha_2^{-1} (M_{\rm str}) &=& \alpha_2^{-1} (m_{\rm SUSY})-
\frac{b_2}{2\pi} \ln \frac{M_{\rm str}}{m_{\rm SUSY}} - \frac{2}{2\pi}
\ln \frac{M_{\rm str}}{M_3},  \\ \alpha_3^{-1} (M_{\rm str})
&=& \alpha_3^{-1} (m_{\rm SUSY}) - \frac{b_3}{2\pi} \ln \frac{M_{\rm
    str}}{m_{\rm SUSY}} - \frac{3}{2\pi} \ln \frac{M_{\rm
    str}}{M_8}, \nonumber 
\end{eqnarray}
where $b_i=(33/5, 1, -3)$ are the coefficients of the one-loop beta
functions of the gauge couplings with MSSM particle content, and
$m_{\rm SUSY}$ is the typical mass scale of the SUSY particles.  To
provide further intuition into the interplay of the messenger scale
and the string scale (where gauge couplings unify), we use
Eq.~(\ref{eq:rgemod}) to write (following the spirit of
\cite{Hisano:1992mh})
\begin{eqnarray} 
(5\alpha_1^{-1} - 3\alpha_2^{-1} -2\alpha_3^{-1})|_{m_{\rm SUSY}} =
  \frac{6}{\pi} \ln \left(\frac{M_{\rm str}^2 M_{\rm mess}}{m_{\rm
      SUSY}^3} \right) \, ,
\end{eqnarray} 
employing a common messenger scale $M_{\rm mess} \equiv M_3 =
M_8$. Putting $\alpha_{1,2,3}^{-1} \simeq (57, 30, 11)$ at $m_{\rm
  SUSY} = 1$ TeV, we obtain
\begin{eqnarray} 
\label{seesaw}
M_{\rm str}^2 M_{\rm mess} = M_G^3 \, .
\end{eqnarray}    
Eq.~(\ref{seesaw}) points to two important things: ($i$) If the
messenger scale lies two orders of magnitude below the GUT scale, then
the scale of gauge unification, which is the string scale, hovers at
one order higher than the GUT scale. ($ii$) Even with the same
supersymmetric mass of $\three$ and $\eight$, i.e. $M_3 = M_8$, the
unification is always maintained. The splitting ($M_3 > M_8$) is
necessarily realized when we require unification at the string scale
by considering two-loop RG running of the gauge couplings. For
instance, taking $M_{\rm str}=5 \times 10^{17}$ GeV and $m_{\rm
  SUSY}=1$ TeV, one obtains $M_3= 1.3 \times 10^{13}$ GeV and $M_8=
3.6 \times 10^{12}$ GeV. In Fig.~{\ref{fig:gut}} we demonstrate this
unification for $M_8 \simeq M_3/6 \simeq 5\times 10^{13} {\rm GeV}$ at
the two-loop level.

\noindent $(ii)$~ The parameters $M_8$ and $\lambda_8 F_X$ control the
squarks and gluino masses, while $M_3$ and $\lambda_3 F_X$ control the
left-slepton and wino masses. Thus the masses of the colored and
uncolored sector become completely independent. Moreover, since
neither $\eight$ nor $\three$ has any non-vanishing hypercharge, the
bino and the right-sleptons are massless at this stage.  A relatively
small mass for them can be induced by gravitational interactions.
Note that this de-correlation of masses has been achieved by the
introduction of separate adjoint messenger multiplets responsible for
the mass generation in the colored and uncolored sectors, {\em
  without} sacrificing the perturbative gauge unification. Moreover,
for the unification to happen at the string scale, one must arrange
$M_3 > M_8$, which helps to keep the left-sleptons lighter than the
squarks.  The elegance of this scenario thus lies in the interlinking
of three issues, namely, perturbative string unification, the presence
of intermediate scales characterizing gauge mediation, and the
relative lightness (more specifically, two orders of magnitude) of
uncolored sparticles compared to the colored ones, including the
extreme lightness of bino and right-sleptons.

Let us now give a closer look into the superparticle spectrum. In our
GMSB setup, the leading contributions to gaugino masses arising from
the messenger loops are given by
\begin{eqnarray}
m_{\tilde{B}} \simeq 0 \, , ~~
m_{\tilde{W}} \simeq \frac{g_2^2}{16\pi^2} (2 \Lambda_3) \, ,~~
m_{\tilde{g}} \simeq  \frac{g_3^2}{16\pi^2} (3 \Lambda_8) \, ,
\end{eqnarray}
where $\Lambda_8 \equiv \lambda_8 F_X /M_8$, $\Lambda_3 \equiv
\lambda_3 F_X /M_3$. Now, considering that $M_3 > M_8$ (discussed in
gauge unification context), we tune $\lambda_8$ and $\lambda_3$ to
ensure $\Lambda_8 \gg \Lambda_3$\footnote{If we impose the universality
  conditions, $M_3=M_8$ and $\lambda_8=\lambda_3$ at $M_{\rm str}$,
  then $\Lambda_8 \simeq \Lambda_3$ holds even at the weak scale since
  $M_8/\lambda_8$ and $M_3/\lambda_3$ are, to a very good
  approximation, RG invariant. However, we do not impose this
  universality as it does not lead us to the condition we require for
  unification, namely, $M_8 \sim M_3/6$.}.  The soft mass-squared
parameters of the squarks and sleptons are given by
\begin{eqnarray}
m_{\tilde{Q}}^2 &\simeq& \frac{2}{(16\pi^2)^2} \left[\frac{4}{3} g_3^4 (3
  \Lambda_8^2) + \frac{3}{4} g_2^4 (2 \Lambda_3^2) \right] \, , ~~
m_{\tilde{D}}^2 = m_{\tilde{U}}^2 \simeq \frac{2}{(16\pi^2)^2}
\frac{4}{3} g_3^4 (3 \Lambda_8^2), \nonumber \\ m_{\tilde{L}}^2 &\simeq&
\frac{2}{(16\pi^2)^2} \frac{3}{4} g_2^4 (2 \Lambda_3^2) \, , ~~
m_{\tilde{E}}^2 \simeq 0.
\end{eqnarray}

For the generation of the bino mass we rely on Planck scale suppressed
gravitational interaction between the supersymmetry breaking field $X$
and the gauge kinetic function
\begin{eqnarray}
\int d^2 \theta \, c_X \frac{X}{M_P} W_\alpha W^\alpha + {\rm h.c.} \,
,
\end{eqnarray}
which also generates wino and gluino masses. The gravitino mass is
generated roughly in the same order (slightly smaller than bino mass)
and is given by
\begin{eqnarray}
m_{3/2} = \frac{F_X}{\sqrt{3} M_P} \, .
\end{eqnarray}
We assume an universal gaugino mass $M_{1/2} \sim c_X 2 \sqrt{3}
m_{3/2} \sim 10 m_{3/2}$ at the unification scale $M_{\rm str}$. The
lightest supersymmetric particle (LSP) turns out to be gravitino,
which is the candidate for dark matter. We also assume a {\em
  sequestered} form of the Kahler potential, for simplicity, to ensure
that tree level contributions to sfermion masses, like $X^\dagger X
\tilde{f}_i^* \tilde{f}_j$, are hugely suppressed. The right-slepton
masses are in fact generated during RG running and are essentially of
the same order as bino mass at the weak scale ($m_{\tilde{E}} \sim 0.4
M_{1/2}$).  Needless to say that the wino and gluino masses pick up
additional messenger induced contributions during the RG running down
to the weak scale.

\begin{figure}[t]
\begin{center}
\includegraphics[scale=1.1]{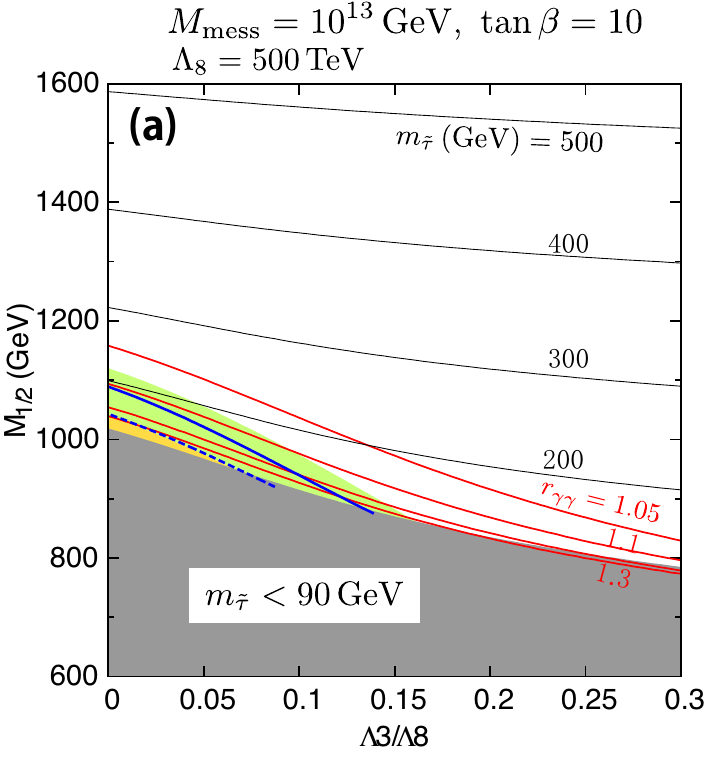}
\includegraphics[scale=1.1]{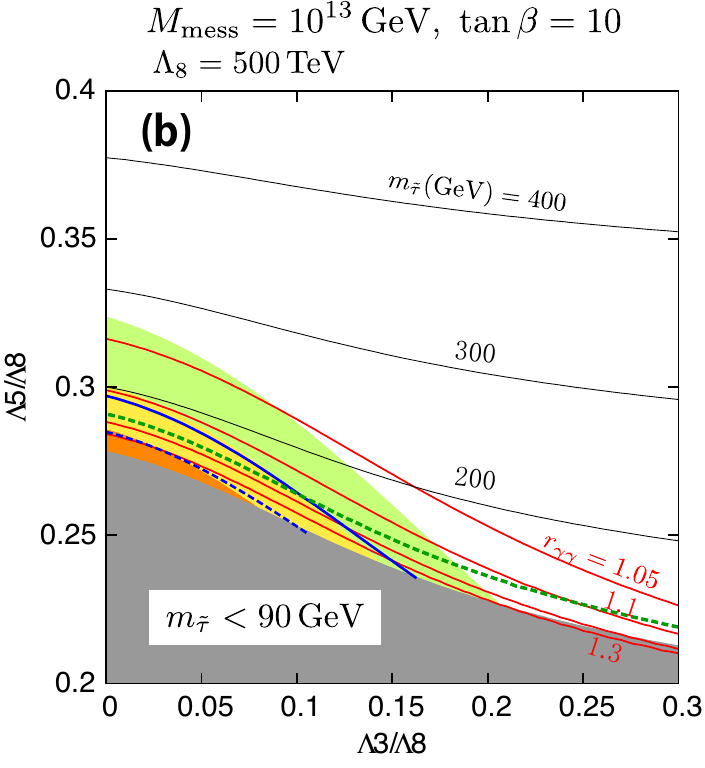}
\caption{\small Contours of diphoton decay rate $r_{\gamma \gamma}
  (=1.05, 1.1, 1.2, 1.3)$ and the (lighter) stau mass. A common
  messenger mass $M_3 = M_8 = M_{\rm mess}=10^{13}$ GeV is assumed,
  for simplicity.  Below the blue solid line the electroweak symmetry
  breaking minimum is unstable (the blue dashed line has been drawn
  with a 10\% relaxation on $\mu\tan\beta$). The large gray shaded
  region at the bottom is disfavored by the experimental lower limit
  on the stau mass. In the {\em left panel (a)}, there is no ${\bf 5 +
    \bar{5}}$ messengers and the stau is the NLSP. Here, in the light
  green (yellow) region, the muon $(g-2)$ is explained at $2
  (1.5)-\sigma$ level. In the {\em right panel (b)}, ${\bf 5 +
    \bar{5}}$ messengers are included to improve consistency with the
  muon $(g-2)$ measurement. In this panel, in the light green
  (yellow) (orange) region, the muon $(g-2)$ is explained at $2 (1.5)
  (1.0)-\sigma$ level. Above (below) the green dashed line the
  neutralino (stau) is the NLSP.}
\label{fig:gmsb}
\end{center}
\end{figure}

In the present scenario, the higgsino mixing parameter $\mu \sim 6$
TeV, which follows from successful electroweak symmetry
breaking. There are two important consequences for such a large $\mu$:
$(i)$ due to a large left-right stau mixing, one of the staus can be
very light, which helps us enhance (though moderately) the diphoton
decay rate of the Higgs; $(ii)$ for the muon $(g-2)$, the bino-smuon
loop numerically dominates over the chargino-sneutrino loop.  In the
{\em left panel (a)} of Fig.~{\ref{fig:gmsb}}, we exhibit the contours
of the lightest stau mass $m_{\tilde{\tau}}$ and the Higgs to diphoton
decay rate normalized to its SM value\footnote{Since squarks,
  including the scalar tops, are super-heavy, the Higgs production
  cross section by gluon fusion is unchanged with respect to the SM.}
$r_{\gamma\gamma} \equiv \Gamma(h\to \gamma \gamma) /\Gamma (h\to
\gamma \gamma)_{\rm SM}$. Note that a (20-25)\% enhancement of Higgs
to diphoton decay rate can be accommodated. In numerical calculations,
we have used the package {\tt SuSpect} \cite{suspect} with appropriate
modifications to include the threshold corrections to the stau and
sleptons from the chargino/neutralino and the heavy Higgs \cite{BPMZ,
  tata}. The muon $(g-2)$ is calculated using {\tt FeynHiggs}
\cite{FeynHiggs}, for which we use the SM prediction as in
\cite{g-2_hagiwara2011}. The region below the blue solid line is
excluded due to the vacuum stability constraint induced by the large
left-right stau mixing \cite{CBV1, CBV2, CBV3}. This constraint puts
an upper limit on $\mu \tan\beta$. In order to impose this constraint,
we have used the fitting formula of \cite{CBV2}. In a more
conservative approach, we have also drawn a blue-dashed line which
corresponds to a relaxation of the upper bound on $\mu\tan\beta$ by
10$\%$.  In the whole region, the stau is the next-to-lightest
supersymmetric particle (NLSP), and the region where
$m_{\tilde{\tau}}$ is below 340 GeV is excluded by the CMS
Collaboration for such a long-lived stau (it can decay into a
gravitino, but the gravitino mass is of the same order as that of the
lighter stau and the coupling is extremely weak)
\cite{long-lived}. But 340 GeV mass of the lighter stau is too heavy
to explain the diphoton enhancement, and it also implies a little
heavier smuon which cannot improve the muon $(g-2)$ discrepancy. The
way out is to admit a mild $R$-parity violation (RPV) which would
allow the stau to promptly decay into a tau (or another lepton) and a
neutrino, thus invalidating the LHC constraint.  Turning on RPV also
invalidates the LHC constraint on the chargino mass
\cite{susy_ew,CMS1}. For convenience, we work only with the trilinear
lepton number violating RPV superpotential as \cite{rpv_reviews}
\begin{eqnarray}
W =  k_1 L L E + k_2  L Q D \, .
\end{eqnarray}
We have omitted the flavor indices for simplicity.  If we assume $k_1,
k_2 \lesssim 10^{-7}$, then baryon asymmetry is not washed out
\cite{washout}. Note that no other constraint is more stringent than
this, and with the couplings of this size, the decay length of stau is
estimated as
\begin{eqnarray}
c \tau_{\tilde{\tau}} = \mathcal{O}(0.1)\, {\rm cm} \left[\frac{(k_1,
    k_2)}{10^{-7}}\right]^{-2} \left( \frac{m_{\tilde{\tau}}}{100\,
  {\rm GeV}}\right)^{-1} \, ,
\end{eqnarray}
which can be regarded as prompt decay. In this case, we can safely
take the stau to be much lighter than 340 GeV, which helps us to enter
within 2-$\sigma$ (at best, 1.5-$\sigma$) allowed zone of muon
$(g-2)$. Note that with this small size of the lepton number violating
couplings, the life-time of the gravitino remains longer than the age
of the universe. Therefore, the gravitino still constitutes a very
good dark matter candidate \cite{Buchmuller:2007ui}.  For the sake of
illustration, we exhibit in Table.~{\ref{table:mass}} the mass
spectrum corresponding to a typical point in the parameter space that
explains the muon $(g-2)$ within 2-$\sigma$.

\begin{table}[t!]
  \begin{center}
    \begin{tabular}{  c | c  }
    $\Lambda_3/\Lambda_8$ & 0.10 \\
    $\Lambda_8$ & 500 TeV \\
    $M_{1/2}$ & 920 GeV \\
    $M_{\rm mess}$ & $10^{13}$ GeV \\
    $\tan\beta$ & 10 \\
    \hline
\hline    
    $\mu$ & $5.9$\,TeV\\
    $m_{\rm stop}$ & 8.2\,TeV \\
    $\delta a_\mu$ & 1.24$\times10^{-9}$ \\
\hline
    $m_{\rm gluino}$ & 10 TeV \\
    $m_{\rm squark}  $ & 9.4 TeV \\
    $m_{\tilde{e}_L} (m_{\tilde{\mu}_L})$ & 601 GeV\\
    $m_{\tilde{e}_R} (m_{\tilde{\mu}_R})$ & 258 GeV \\
    $m_{\tilde{\tau}_1}$ & 98 GeV \\
    $m_{\chi_1^0}$ & 315 GeV \\
     $m_{\chi_1^{\pm}}$ & 851 GeV \\
    \end{tabular}
        \begin{tabular}{  c | c  }
    $\Lambda_3/\Lambda_8$ & 0.10 \\
    $\Lambda_5/\Lambda_8$ & 0.28 \\
    $\Lambda_8$ & 500 TeV \\
    $M_{\rm mess}$ & $10^{13}$ GeV \\
    $\tan\beta$ & 10 \\
    \hline
\hline    
    $\mu$ & $5.6$\,TeV\\
    $m_{\rm stop}$ & 7.8\,TeV \\
    $\delta a_\mu$ & 1.12 $\times10^{-9}$ \\
\hline
    $m_{\rm gluino}$ & 9.5 TeV \\
    $m_{\rm squark}  $ & 8.9 TeV \\
    $m_{\tilde{e}_L} (m_{\tilde{\mu}_L})$ & 574 GeV\\
    $m_{\tilde{e}_R} (m_{\tilde{\mu}_R})$ & 310 GeV \\
    $m_{\tilde{\tau}_1}$ & 201 GeV \\
    $m_{\chi_1^0}$ & 156 GeV \\
     $m_{\chi_1^{\pm}}$ & 557 GeV \\
    \end{tabular}
        \begin{tabular}{  c | c  }
    $\Lambda_3/\Lambda_8$ & 0.05 \\
    $\Lambda_5/\Lambda_8$  & 0.27 \\
    $\Lambda_8$ & 500 TeV \\
    $M_{\rm mess}$ & $10^{13}$ GeV \\
    $\tan\beta$ & 10 \\
    \hline
\hline    
    $\mu$ & $5.6$\,TeV\\
    $m_{\rm stop}$ & 7.8\,TeV \\
    $\delta a_\mu$ & 1.86 $\times10^{-9}$ \\
\hline
    $m_{\rm gluino}$ & 9.5 TeV \\
    $m_{\rm squark}  $ & 8.9 TeV \\
    $m_{\tilde{e}_L} (m_{\tilde{\mu}_L})$ & 451 GeV\\
    $m_{\tilde{e}_R} (m_{\tilde{\mu}_R})$ & 289 GeV \\
    $m_{\tilde{\tau}_1}$ & 105 GeV \\
    $m_{\chi_1^0}$ & 149 GeV \\
     $m_{\chi_1^{\pm}}$ & 411 GeV \\
    \end{tabular}
    \caption{\small Sample mass spectra shown in three vertical
      blocks. The left block corresponds to the case without the
      conventional ${\bf 5 (\bar{5})}$ messengers, where the stau is
      the NLSP. The right two blocks correspond to the cases including
      the ${\bf 5 (\bar{5})}$ messengers. In the second block the
      lightest neutralino is the NLSP while in the third block the
      stau is the NLSP.}
  \label{table:mass}
  \end{center}
\end{table}

How can we enter inside the 1-$\sigma$ allowed zone of muon $(g-2)$?
For that we need to give up the minimality of messenger particle
content, and add a $\five$ and a $\fivebar$ to the existing $\three$
and $\eight$, i.e. add to Eq. (\ref{eq:mess1}) the following piece
\cite{yanagida-old}
\begin{eqnarray}
W_{\rm mess}^{\rm new}= (M_5 +  \lambda_5 X ) \five \fivebar \, .
\end{eqnarray}
This of course does not alter the scale of string unification.  But
what do we gain by this? In this scenario, bino and right-slepton
masses are generated by $\five$ and $\fivebar$ as they have
non-vanishing hypercharges (as in conventional GMSB).  We can
completely ignore the supergravity effects for their mass
generation. Importantly, we do not need to make any {\em ad hoc}
assumption regarding the sequestering of Kahler potential.  Now we
have the freedom of choosing either bino or stau as NLSP (unlike in
the previous scenario where stau is necessarily the NLSP). Gravitino
is as usual the LSP but it can be much lighter, e.g. $\sim$ 1 GeV,
than in the previous case ($\sim$ 10 GeV).  In the {\em right panel
  (b)} of Fig.~{\ref{fig:gmsb}}, we show that contours of
$m_{\tilde{\tau}}$, $r_{\gamma \gamma}$ and the muon $(g-2)$.  A
common messenger scale $M_{\rm mess}=10^{13}$ GeV has been
taken. Above the green dashed line, the neutralino is the NLSP. But if
$R$-parity is conserved, the neutralino NLSP is sufficiently
long-lived to contradict bounds from big-bang nucleosynthesis, to
avoid which we need a mild RPV interaction.  Below the green dashed
line the stau is the NLSP, and for such a light stau one again
requires a mild RPV interaction. But the price of turning on RPV is
compensated as in this region the muon $g-2$ can be explained at
1-$\sigma$ level.  In the second and third vertical blocks of
Table~{\ref{table:mass}} we have shown mass spectra of typical points
for the neutralino NLSP and stau NLSP scenarios, respectively.

A word of caution in now in order.  Since the different gauginos
originate from independent messenger multiplets, their masses may pick
up additional independent phases.  These phases along with the one
associated with the supersymmetry breaking $B$-parameter of the scalar
potential need to be appropriately tuned, especially when the sleptons
and the weak gauginos are very light for accommodating the muon
$(g-2)$, to ensure that the tension with the electron electric dipole
moment is not further aggravated. The remedy may lie in choosing some
specific ultraviolet completion which might align these phases.
Alternatively, one may invoke some symmetry that maintains
CP-invariance in the hidden sector, or make an assumption that weak
scale CP-violation manifests only through the CKM matrix. A more
elaborate discussion on this is beyond the scope of this paper.

Our scenario predicts the presence of light sleptons and electroweak
gauginos which may be observed in collider experiments. In most of the
parameter space of our interest, $\tilde{\tau}_1$ is the NLSP and the
decay modes of $\tilde{\tau}_1$ can be classified into three
categories, e.g., $\tilde{\tau}_1 \rightarrow e/\mu + \nu$,
$\tilde{\tau}_1 \rightarrow \tau + \nu$ and $\tilde{\tau}_1
\rightarrow q q^{'}$, depending on the particular form of the RPV
operator. Electroweak gaugino production is not negligible and
gauginos decay mainly to $\tilde{\tau}_1$. This leads to final states
with lepton(s) and missing transverse energy. The CMS collaboration
has searched for electroweak gauginos and sleptons in multi-lepton
(including $\tau$) plus missing transverse energy final states
\cite{CMS1}. So far, the bounds are not strong enough to exclude our
scenario. For example, the lower limit on the chargino mass is about
300 GeV if $\tilde{\tau}_1$ decays exclusively to $\tau$. It is
expected that 14 TeV LHC can cover most of the parameter space which
is consistent with the result of muon $(g-2)$ experiment. However,
direct pair production of $\tilde{\tau}_1$ (which is mostly the
right-type) and its decay to jets is difficult to be searches at the
LHC. Even in that case, $\tilde{\tau}_1$ with the mass up to 250-500
GeV can be discovered at the proposed ILC with center of mass energy
(0.5-1.0) TeV.

{\em In conclusion}, the present experimental context, especially the
discovery of a 125 GeV Higgs-like boson at the LHC, and
non-observation of anything else {\em new} either at the LHC or in
other experiments, compels us to have a sincere introspection of many
a scenarios beyond the SM that we have so far been advocating. What
about supersymmetry? Does naturalness demand that all superparticles
have to be simultaneously heavy? Or, can we still have room for some
light superpartners, e.g. sleptons and weak gauginos, with two orders
of magnitude mass splitting between them and the squarks and gluino
created by a natural dynamics?  This is precisely the question we have
asked in this paper.  Our key observation is that by employing an {\em
  unconventional} (i.e. not the conventional ${\bf 5}$-plets) choice
of messenger multiplets, namely, a color SU(3) octet and a weak SU(2)
triplet with $M_3 > M_8$ in a GMSB setup, we can generate a spectrum
that naturally accommodates the required mass splitting between the
colored and uncolored superpartners, justifying all current data
including the muon $(g-2)$ and also a very moderate enhancement in the
Higgs decay width in diphoton mode.  What is really elegant about this
scenario is that the use of even {\em incomplete} SU(5) messenger
multiplets at an intermediate scale does not disturb a successful
unification of the gauge couplings. Depending on the choice of the
triplet and the octet messenger masses (still maintaining $M_3 >
M_8$), the meeting point moves to a scale slightly higher than that of
usual grand unification; and interestingly, the new scale can, in
fact, be the string scale where the gauge and gravitational couplings
are perturbatively unified \cite{yanagida-original}. Thus,
perturbative string unification, intermediate scales characterizing
GMSB messengers, and the relative lightness of sleptons and weak
gauginos compared to the squarks and the gluino, are all interlinked
by a single underlying dynamics. This scenario is testable, as these
uncolored superpartners can be discovered (or excluded) at the 14 TeV
LHC or at the ILC.

\noindent {\bf Acknowledgements:}~ G.B. thanks Kavli IPMU for
hospitality when the work was done. N.Y. thanks S. Sugimoto for useful
discussion.  We all thank M. Ibe and S. Matsumoto for discussions at the
early stage of this work.  The work of N.Y. is supported in part by
JSPS Research Fellowships for Young Scientists.  This work is also
supported by the World Premier International Research Center
Initiative (WPI Initiative), MEXT, Japan.



\begin{thebibliography}{99}
\bibitem{Aad:2012tfa} 
  G.~Aad {\it et al.}  [ATLAS Collaboration],
  Phys.\ Lett.\ B {\bf 716}, 1 (2012)
  [arXiv:1207.7214 [hep-ex]].

\bibitem{Chatrchyan:2012ufa} 
  S.~Chatrchyan {\it et al.}  [CMS Collaboration],
  Phys.\ Lett.\ B {\bf 716}, 30 (2012)
  [arXiv:1207.7235 [hep-ex]].


\bibitem{OYY}
  Y.~Okada, M.~Yamaguchi and T.~Yanagida,
  Prog.\ Theor.\ Phys.\ {\bf 85}, 1
  (1991); J.~R.~Ellis, G.~Ridolfi and F.~Zwirner,
  Phys.\ Lett.\ B {\bf 257}, 83 (1991);
  H.~E.~Haber and R.~Hempfling,
  Phys.\ Rev.\ Lett.\  {\bf 66}, 1815 (1991);
  J.~R.~Ellis, G.~Ridolfi and F.~Zwirner,
  Phys.\ Lett.\ B {\bf 262}, 477 (1991).

\bibitem{Okada:1990gg} 
  Y.~Okada, M.~Yamaguchi and T.~Yanagida,
  Phys.\ Lett.\ B {\bf 262}, 54 (1991).


\bibitem{susy_sq_glu}
  [ATLAS Collaboration],
  ATLAS-CONF-2012-109;
  S.~Chatrchyan {\it et al.}  [CMS Collaboration],
  arXiv:1301.2175 [hep-ex].

\bibitem{susy_ew}
  G.~Aad {\it et al.}  [ATLAS Collaboration],
  Phys.\ Lett.\ B {\bf 718}, 879 (2013)
  [arXiv:1208.2884 [hep-ex]].

\bibitem{CMS1}
The CMS Collaboration, CMS-PAS-SUS-12-022.


\bibitem{muon_g_exp} 
  G.~W.~Bennett {\it et al.}  [Muon G-2 Collaboration],
  Phys.\ Rev.\ D {\bf 73}, 072003 (2006)
  [hep-ex/0602035].


\bibitem{g-2_hagiwara2011}
  K.~Hagiwara, R.~Liao, A.~D.~Martin, D.~Nomura, and T.~Teubner,
  J.\ Phys.\ G {\bf 38}, 085003 (2011)
  [arXiv:1105.3149 [hep-ph]].


\bibitem{g-2_davier2010}
  M.~Davier, A.~Hoecker, B.~Malaescu, and Z.~Zhang,
  Eur.\ Phys.\ J.\ C {\bf 71}, 1515 (2011)
  [Erratum-ibid.\ C {\bf 72}, 1874 (2012)]
  [arXiv:1010.4180 [hep-ph]].

\bibitem{susy_muon_g}
  U.~Chattopadhyay and P.~Nath,
  Phys.\ Rev.\ D {\bf 53}, 1648 (1996)
  [hep-ph/9507386];
  T.~Moroi,
  Phys.\ Rev.\ D {\bf 53}, 6565 (1996)
  [Erratum-ibid.\ D {\bf 56}, 4424 (1997)]
  [hep-ph/9512396];
  M.~S.~Carena, G.~F.~Giudice and C.~E.~M.~Wagner,
  Phys.\ Lett.\ B {\bf 390}, 234 (1997)
  [hep-ph/9610233];
  S.~P.~Martin and J.~D.~Wells,
  Phys.\ Rev.\ D {\bf 64}, 035003 (2001)
  [hep-ph/0103067];
  J.~L.~Feng and K.~T.~Matchev,
  Phys.\ Rev.\ Lett.\  {\bf 86}, 3480 (2001)
  [hep-ph/0102146].




\bibitem{atlas_diphoton} Fabrice Hubaut, ATLAS Collaboration,
  \href{https://indico.in2p3.fr/getFile.py/access?contribId=45&sessionId=6&resId=0&materialId=slides&confId=7411}{Talk
    at the Moriond 2013 EW session}. Eleni Mountricha, ATLAS
  Collaboration,
  \href{http://moriond.in2p3.fr/QCD/2013/ThursdayMorning/Mountricha.pdf}{Talk
    at the Moriond 2013 QCD session}.

\bibitem{Carena:2012gp} 
  M.~Carena, S.~Gori, N.~R.~Shah, C.~E.~M.~Wagner and L.~-T.~Wang,
  JHEP {\bf 1207}, 175 (2012)
  [arXiv:1205.5842 [hep-ph]].

\bibitem{Cao:2012fz} 
  J.~-J.~Cao, Z.~-X.~Heng, J.~M.~Yang, Y.~-M.~Zhang and J.~-Y.~Zhu,
  JHEP {\bf 1203}, 086 (2012)
  [arXiv:1202.5821 [hep-ph]].

\bibitem{Ajaib:2012eb} 
  M.~A.~Ajaib, I.~Gogoladze and Q.~Shafi,
  Phys.\ Rev.\ D {\bf 86}, 095028 (2012)
  [arXiv:1207.7068 [hep-ph]].



\bibitem{cms_diphoton} Christophe Ochando, CMS collaboration,
  \href{http://moriond.in2p3.fr/QCD/2013/ThursdayMorning/Ochando.pdf}{Talk
    at the Moriond 2013 QCD session}.



\bibitem{GMSB} 
  M.~Dine and A.~E.~Nelson,
  Phys.\ Rev.\ D {\bf 48}, 1277 (1993)
  [hep-ph/9303230];
  M.~Dine, A.~E.~Nelson and Y.~Shirman,
  Phys.\ Rev.\ D {\bf 51}, 1362 (1995)
  [hep-ph/9408384];
  M.~Dine, A.~E.~Nelson, Y.~Nir and Y.~Shirman,
  Phys.\ Rev.\ D {\bf 53}, 2658 (1996)
  [hep-ph/9507378].

 
\bibitem{yanagida-old} 
 T.~Han, T.~Yanagida and R.~-J.~Zhang,
  Phys.\ Rev.\ D {\bf 58}, 095011 (1998)
  [hep-ph/9804228].
  
\bibitem{yanagida-recent}
M.~Ibe, S.~Matsumoto, T.~T.~Yanagida and N.~Yokozaki,
  arXiv:1210.3122 [hep-ph].

\bibitem{yanagida-original} 
  C.~Bachas, C.~Fabre and T.~Yanagida,
  Phys.\ Lett.\ B {\bf 370}, 49 (1996)
  [hep-th/9510094].


\bibitem{string} 
  K.~R.~Dienes,
  Phys.\ Rept.\  {\bf 287}, 447 (1997)
  [hep-th/9602045].

\bibitem{Hisano:1992mh} 
  J.~Hisano, H.~Murayama and T.~Yanagida,
  Phys.\ Rev.\ Lett.\  {\bf 69}, 1014 (1992).

\bibitem{suspect} 
  A.~Djouadi, J.~-L.~Kneur and G.~Moultaka,
  Comput.\ Phys.\ Commun.\  {\bf 176}, 426 (2007)
  [hep-ph/0211331].


\bibitem{BPMZ} 
  D.~M.~Pierce, J.~A.~Bagger, K.~T.~Matchev and R.~-j.~Zhang,
  Nucl.\ Phys.\ B {\bf 491}, 3 (1997)
  [hep-ph/9606211].
 
\bibitem{tata} 
  A.~D.~Box and X.~Tata,
  Phys.\ Rev.\ D {\bf 79}, 035004 (2009)
  [Erratum-ibid.\ D {\bf 82}, 119905 (2010)]
  [arXiv:0810.5765 [hep-ph]].


\bibitem{FeynHiggs} 
  S.~Heinemeyer, W.~Hollik and G.~Weiglein,
  Comput.\ Phys.\ Commun.\ \ {\bf 124}, 76  (2000)
  [hep-ph/9812320];
  S.~Heinemeyer, W.~Hollik and G.~Weiglein,
  Eur.\ Phys.\ J.\ C\ {\bf 9}, 343  (1999)
  [hep-ph/9812472];
  G.~Degrassi, S.~Heinemeyer, W.~Hollik, P.~Slavich and G.~Weiglein,
  Eur.\ Phys.\ J.\ C\ {\bf 28}, 133  (2003)
  [hep-ph/0212020].
  M.~Frank, T.~Hahn, S.~Heinemeyer, W.~Hollik, H.~Rzehak and G.~Weiglein,
  JHEP\ {\bf 0702}, 047  (2007)
  [hep-ph/0611326].


 
 \bibitem{CBV1} 
  R.~Rattazzi and U.~Sarid,
  Nucl.\ Phys.\ B {\bf 501}, 297 (1997)
  [hep-ph/9612464].

\bibitem{CBV2} 
  J.~Hisano and S.~Sugiyama,
  Phys.\ Lett.\ B {\bf 696}, 92 (2011)
  [Erratum-ibid.\ B {\bf 719}, 472 (2013)]
  [arXiv:1011.0260 [hep-ph]].

 \bibitem{CBV3}
   M.~Carena, S.~Gori, I.~Low, N.~R.~Shah and C.~E.~M.~Wagner,
  JHEP {\bf 1302}, 114 (2013)
  [arXiv:1211.6136 [hep-ph]];
 T.~Kitahara and T.~Yoshinaga,
  arXiv:1303.0461 [hep-ph].



\bibitem{long-lived}
The CMS Collaboration, CMS PAS EXO-12-026.

\bibitem{rpv_reviews}
For reviews see, 
  G.~Bhattacharyya,
  Nucl.\ Phys.\ Proc.\ Suppl.\  {\bf 52A}, 83 (1997) 
  [arXiv:hep-ph/9608415];
G.~Bhattacharyya,
  arXiv:hep-ph/9709395;
H.~K.~Dreiner,
  arXiv:hep-ph/9707435;
M.~Chemtob,
  Prog.\ Part.\ Nucl.\ Phys.\  {\bf 54}, 71 (2005)
  [arXiv:hep-ph/0406029];
  R.~Barbier {\it et al.},
  Phys.\ Rept.\  {\bf 420}, 1 (2005) 
  [arXiv:hep-ph/0406039];
Y.~Kao and T.~Takeuchi,
  arXiv:0910.4980 [hep-ph].



\bibitem{washout}
  B.~A.~Campbell, S.~Davidson, J.~R.~Ellis and K.~A.~Olive,
  Phys.\ Lett.\  B {\bf 256}, 457 (1991); 
  W.~Fischler, G.~F.~Giudice, R.~G.~Leigh and S.~Paban,
  Phys.\ Lett.\  B {\bf 258}, 45 (1991); 
  H.~K.~Dreiner and G.~G.~Ross,
  Nucl.\ Phys.\  B {\bf 410}, 188 (1993) 
  [arXiv:hep-ph/9207221].


\bibitem{Buchmuller:2007ui} 
  W.~Buchmuller, L.~Covi, K.~Hamaguchi, A.~Ibarra and T.~Yanagida,
  JHEP {\bf 0703}, 037 (2007)
  [hep-ph/0702184 [HEP-PH]], and references therein.


\end{thebibliography}
\end{document}